%% file: scbosonization.tex
\documentclass[11pt,a4paper]{article}
\usepackage{jheppub}

\usepackage{amssymb,amsmath}
\usepackage{amsfonts,amsthm}
\usepackage{youngtab}
\usepackage{tikz}
\usepackage{subcaption}

\usetikzlibrary{shapes.misc,mindmap,backgrounds,decorations.pathreplacing,decorations.markings}
\usetikzlibrary{decorations.pathreplacing,decorations.pathmorphing,calc}
\newcommand{\tikzmark}[1]{\tikz[overlay,remember picture] \node (#1) {};}

\title{Bosonization in Non-Relativistic CFTs}
\author[a]{Carl Turner}
\affiliation[a]{Department of Applied Mathematics and Theoretical Physics, \\
University of Cambridge, \\ 
Cambridge, CB3 OWA, UK}
\emailAdd{C.P.Turner@damtp.cam.ac.uk}
\abstract{We demonstrate explicitly the correspondence between all protected operators in a 2+1 dimensional non-supersymmetric bosonization duality in the non-relativistic limit. Roughly speaking we consider $SU(N)$ Chern-Simons field theory at level $k$ with $N_f$ flavours of fundamental boson, and match its chiral sector to that of a $SU(k)$ theory at level $N$ with $N_f$ fundamental fermions. We present the matching at the level of indices and individual operators, seeing the mechanism of failure for $N_f > N$, and point out that the non-relativistic setting is a particularly friendly setting for studying interesting questions about such dualities.}

\def\cD{{\cal D}}
\def\cO{{\cal O}}
\def\cJ{{\cal J}}
\def\rmd{{\rm d}}

\def\tr{\operatorname{tr}\,}
\def\sign{\operatorname{sgn}}

\def\fstp{\,\mbox{.}}

\makeatletter

\newcommand\cpttikzundefinenode[1]{
  \expandafter\ifx\csname pgf@sh@ns@#1\endcsname\relax
  \else
    \typeout{===>Undefining node "#1"}
    \global\expandafter\let\csname pgf@sh@ns@#1\endcsname\relax
  \fi
}
\newcommand\cpttikzundefinethesenodes[1]{
  \foreach \myn  in {#1}
    {
      \expandafter\cpttikzundefinenode\expandafter{\myn}
    }
}

\makeatother

\def\ktrue{k}
\def\ksh{{\hat{k}}}
\def\rnk{N}

\def\lag{\mathcal{L}}

\begin{document}

\maketitle


\clearpage
\section{Introduction}

One of the highlights of recent work in the theoretical physics community has been a significant leap in our understanding of field theoretic dualities, especially in 3 dimensions. Part of this is a new appreciation for old ideas from condensed matter physics suggesting relations between theories of bosons and fermions \cite{polyakov,fradhap,shankar1,shankar2} and particle-vortex duality \cite{peskin,dh} -- and newer ideas about fermionic dualities \cite{son,senthil} -- and part of it has its origins in large $N$ physics \cite{guy1,min1,ofer1}. These have been united \cite{shiraz4,ofer,hsin} (with supersymmetry \cite{givkut,benini,ofer3d4d,parksandrecreation,min2,guy3} as a guiding principle, and level-rank duality \cite{levelrank} as the algebraic heart) into a coherent set of precise statements about the equality of partition functions for (roughly speaking) $SU(N)_k$ bosonic theories and $SU(k)_N$ fermionic theories, forming an elaborate web of dualities \cite{karchtong,ssww}.

Meanwhile, in \cite{cpt-anyons,cpt-nonanyon}, the spectrum of a broad class of \textit{non-relativistic} Chern-Simons CFTs was analyzed. In particular, by exploiting a close relationship with a superconformal theory, the exact dimensions of a set of protected operators were calculated. 

But, as discussed in \cite{cpt-dualityqhe}, these non-relativistic CFTs inherit certain dualities from their relativistic parents. In particular, the \emph{bosonization} dualities proposed in \cite{ofer} (and generalized slightly in \cite{cpt-dualityqhe}) have well-defined non-relativistic limits offering exact correspondences between non-relativistic Chern-Simons-matter theories coupled to bosons and fermions. Of course, non-relativistic field theory is essentially quantum mechanics, and these dualities are fancier versions of flux attachment \cite{frank}. Yet the nature of bosonization remains non-trivial even in this simpler setting.

In fact, the chiral physics we discuss is intimately related to the WZW model -- indeed, see \cite{cpt-nawzw} for a direct map between a different phase of this theory and the WZW model -- and the 1+1 dimensional non-Abelian bosonization story \cite{Witten-nonab}.

In this paper, we will explain how bosonization is realized in the setting of non-relativistic Chern-Simons-matter field theories. We begin in this introduction with an overview of the results of this paper. Then in section \ref{definition-sec} we give clear statements of the pairs of theories which are dual in this sense, and recapitulate the relevant results of \cite{cpt-anyons,cpt-nonanyon}. This done, in section \ref{sec-protected} we will present the correspondence between the protected states whose dimensions are known exactly in these theories. In section \ref{analysis-sec} we will discuss this matching and its limitations. Finally, in section \ref{conclusion} we will conclude with a brief summary.

\subsection*{Key Results}

The bosonization dualities discussed above generically look like a $SU(\rnk)_k \leftrightarrow SU(k)_{\rnk}$ level-rank duality with some extra structure. We will demonstrate explicitly the non-supersymmetric duality between the following two Chern-Simons-matter theories:

\begin{samepage}
\paragraph{Bosonic Theory:} $U(\rnk)_{\ktrue,\ktrue+n\rnk}$ coupled to $N_f$ fundamental non-relativistic scalars.
\nopagebreak
\paragraph{Fermionic Theory:} $U(k)_{-\rnk,-\rnk}$ coupled to $N_f$ fundamental non-relativistic fermions and, through a BF coupling, to $U(1)_n$.
\end{samepage}

We will explain various subtleties here (especially regarding the $U(1)$ parts of this duality) in section \ref{definition-sec}. For now, we merely remark that the first and second subscripts on each gauge group indicate the non-Abelian and Abelian Chern-Simons levels respectively. Importantly, we will find that the duality holds only for $N_f \le N$.

Both theories will exhibit non-relativistic conformal symmetry \cite{aff,roman,nishidason}, associated to an extra $SO(2,1)$ symmetry beyond the naive Galilean group. This is very analogous to the more familiar relativistic conformal symmetry, with features such as a state-operator map, though it is generally simpler to understand. In particular, in previous work the dimensions of a protected class of \textit{chiral} operators were calculated exactly \cite{cpt-anyons,cpt-nonanyon}.

We will find that we can give a precise mapping between the chiral operators of both theories. As we shall see and explain, this even includes operators transforming non-trivially under the global part of the gauge group. The mapping takes the form of a simple explicit recipe to construct the dual of any given operator. There is a close relation to the world of 1+1 dimensional level-rank duality of conformal field theory, which we will briefly outline.

It can also, of course, be encoded as an equality of certain indices between the two theories. We will describe how to construct these indices by starting with the spectrum of free particles and implementing the flux attachment. The mapping and indices are given in section \ref{sec-protected}.

In order to make this duality fly, we have to understand an important subtlety of Chern-Simons theory: not all operators one might naively write down are permitted in the spectrum. This is dictated by the \textit{fusion rules} of the underlying affine Lie algebra. Normally, when one brings together $p$ particles transforming in representations $R_i$ of some group, they collectively transform in arbitrary irreps in the tensor product $R_1 \otimes R_2 \otimes \cdots \otimes R_p$. However, the fusion rules knock out certain terms in that decomposition into irreps. In particular, in the language of Young diagrams, no representation of $SU(N)_k$ with more than $k$ columns is permitted.

Understanding precisely how these are implemented has important, and slightly subtle, ramifications for our spectra -- we will address this fully only in section
\ref{analysis-sec}
as we go through several simple examples of the duality, understanding how fusion rules are implemented and then investigating how the number of flavours affects the duality. We will find that the duality stated holds only for $N_f \le N$, as the fermionic theory given above contains extra states.

One nice perspective on this duality is that it is a return to the world bosonization originally came from: non-relativistic flux attachment. When we are done, we will have a detailed and very down-to-earth understanding of exactly how attaching flux to bosons and fermions works (including how this is restricted by fusion rules, and what it looks like at the level of partition functions or indices) and exactly how this gives rise to dualities.

Such a simple framework makes for an excellent testbed for investigating more potential dualities, and we leave this open for further work.


\clearpage
\section{Duality of Non-Relativistic Conformal Theories}\label{definition-sec}

Non-relativistic theories, and especially those with conformal symmetry, are not particularly familiar to many, so in this section we will summarize some of their key properties. We will begin by describing their Lagrangians, before taking a moment to discuss some of the subtleties of the duality as relevant for these theories. Then we will recapitulate results about their spectra. This will leave us ready to begin explaining how the duality is implemented in section \ref{sec-protected}.

\subsection{Lagrangians}

Recall that we are interested in the duality
\begin{equation}\label{duality}
U(\rnk)_{\ktrue,\ktrue+n\rnk} + N_f \text{ scalars} \qquad \longleftrightarrow \qquad U(k)_{-\rnk,-\rnk} \times U(1)_n + N_f \text{ fermions}
\end{equation}
where both theories are non-relativistic and have conformal symmetry.

In order to specify these theories completely, we specify their gauge and matter sectors separately. Firstly, we should explain the notation $U(\rnk)_{\kappa,\kappa'}$. This refers to the gauge group
\begin{equation}\label{upkkp}
U(\rnk)_{\kappa,\kappa'} = \frac{ U(1)_{\kappa' \rnk} \times SU(\rnk)_{\kappa} } {\mathbb{Z}_\rnk}
\end{equation}
with the corresponding Lagrangian
\begin{equation*}
\lag_{\rm CS} = -\frac{\kappa}{4\pi} \tr \epsilon^{\mu\nu\rho} \left(a_\mu \partial_\nu a_\rho - \frac{2i}{3} a_\mu a_\nu a_\rho\right) - \frac{\kappa' \rnk}{4\pi} \epsilon^{\mu\nu\rho} \tilde{a}_\mu \partial_\nu \tilde{a}_\rho
\end{equation*}
where $a$ is the $SU(\rnk)$ part of the gauge field and $\tilde{a}$ is the $U(1)$ part. Notice that the discrete quotient of \eqref{upkkp} enforces the constraint $\kappa'-\kappa \in \rnk\mathbb{Z}$. Indeed, the bosonic theory exhibits the general solution of this constraint, using the parameter $n$.

For the particular values of $n=0,\pm 1,\infty$ it reduces to special cases analyzed in \cite{ofer,hsin} which are dual to fermionic theories with a single gauge field. These are realized as follows:
\begin{align*}
&n=1: & U(N)_{k,k+N} + N_f \text{ scalars} &\longleftrightarrow U(k)_{-N,-N-k} + N_f \text{ fermions} \\
&n=-1: & U(N)_{k,k-N} + N_f \text{ scalars} &\longleftrightarrow U(k)_{-N,-N+k} + N_f \text{ fermions} \\
&n=0: & U(N)_{k,k} + N_f \text{ scalars} &\longleftrightarrow SU(k)_{-N} + N_f \text{ fermions} \\
&n=\infty: & SU(N)_{k} + N_f \text{ scalars} &\longleftrightarrow U(k)_{-N,-N} + N_f \text{ fermions}
\end{align*}
For general $n$, the fermionic theory requires an extra $U(1)_n$ factor which strictly should not be integrated out. This is coupled in via a so-called BF term, which couples the overall $U(1) \subset U(k)$ field $\tilde{a}$ to the extra $U(1)_n$ field $b$ via the simple term
\begin{equation*}
\lag_{\rm BF} = \frac{k}{2\pi} \epsilon^{\mu\nu\rho} \tilde{a}_\mu \partial_\nu b_\rho \fstp
\end{equation*}
Nonetheless, these are all easily related to each other by gauging $U(1)$ symmetries and integrating out fields, so in the following we will mainly find it simplest to work with the $n=1$ case in which the Abelian and (renormalized) non-Abelian levels come out to be the same.

All that remains is to describe the matter couplings. These are simply given by
\begin{align*}
\lag_{\rm bosonic} &= i \phi^\dagger_i \cD_0 \phi_i - \frac{1}{2m} \vec{\cD} \phi^\dagger_i \cdot \vec{\cD} \phi_i - \frac{1}{2m} \phi^\dagger_i f_{12} \phi_i \\
\lag_{\rm fermionic} &= i \psi^\dagger_i \cD_0 \psi_i - \frac{1}{2m} \vec{\cD} \psi^\dagger_i \cdot \vec{\cD} \psi_i - \frac{1}{2m} \psi^\dagger_i f_{12} \psi_i
\end{align*}
where $i = 1,\ldots,N_f$ is a flavour index, $\cD = d-ia-i\tilde{a}$ denotes the appropriate covariant derivative, and $f$ is the field strength associated to the connection $a+\tilde{a}$. We have suppressed gauge indices.\footnote{Note that we make a different sign choice for the last Pauli-type term to that originally made in \cite{cpt-anyons}. This is to ensure that we get the theory with the correct parity.}

It is helpful to see the form of Gauss's laws, the equations of motion for $a_0$ and $\tilde{a}_0$, for such a theory. Looking at just the Abelian sector of the bosonic theory for an example, we find
\begin{equation}\label{gauss-bose}
\qquad \tr f_{12} = \frac{2\pi}{k+n\rnk} \phi^\dagger_i \phi_i \fstp
\end{equation}
Hence the last term in each action is essentially a quartic self-interaction (equivalently, a two-body delta-function interaction) for the matter fields with a very special value. As discussed in \cite{cpt-anyons,cpt-nonanyon} following \cite{bergman,bbak}, this is a natural choice in a renormalization group sense (not too surprising given the theory is conformal here). It is also special in that there is a supersymmetric completion of such Lagrangians; similarly, this is the Bogomolny point.

This is all reflected in the fact that one can rewrite the above terms in a very neat way, leading to the Hamiltonians
\begin{equation}\label{hams}
\begin{aligned}
H_{\rm bosonic} &= \int \rmd^2x \ \frac{2}{m} \left| \cD_z \phi_i \right|^2 \\
H_{\rm fermionic} &= \int \rmd^2x \ \frac{2}{m} \left| \cD_z \psi_i \right|^2 \fstp
\end{aligned}
\end{equation}

\subsection{Subtleties of the Duality}

Bosonization dualities in non-Abelian theories are essentially level-rank dualities of the underlying gauge groups, so we expect that, loosely speaking, a $SU(\rnk)_k$ theory with bosons should be dual to a $SU(k)_{\rnk}$ theory with fermions. However, there are several subtleties which have been addressed to arrive at \eqref{duality}:

\begin{enumerate}
\item The level-rank duality does not deal with the $U(1)$ parts of the dual theories. In fact, as first clarified in \cite{ofer}, one always needs to include non-trivial $U(1)$ gauge groups in implementing these dualities.
\item It is possible that the implementation of bosonization may involve a discrete symmetry: parity. This is in fact the case -- indeed, the $SU(\rnk)_k$ theory is dual to something like $SU(k)_{-\rnk}$ (see, for instance, \cite{hsin}).
\item With the introduction of $N_f$ fermionic fields on one side of the duality, there is a renormalization of the Chern-Simons level. In the relativistic case, this shift is by $\pm N_f/2$ according to other sign choices. (One can think of this as arising from integrating out half of the fermions in the analogous supersymmetric duality, mirror symmetry \cite{Kachru2016}.)
\item In taking the non-relativistic limit of a matter field theory, one gives all fields a mass and then decouples either the particles or anti-particles. Consequently, there is a further shift of $\pm N_f/2$ in the fermionic theories according to what choice one makes here. For the theories discussed in this paper, this shift precisely negates the previous shift.\footnote{In \cite{cpt-dualityqhe}, these two contributions added together. The reason for the distinction is that we make a different choice of how to integrate out the fermions. Presumably this is a distinction associated to the choice of ground state of the theory; here we are interested in the conformal phase.}
\item Finally, there is a further possible renormalization of the non-Abelian Chern-Simons level according to the choice of regularization for loop integrals in the field theory. This comes from gluon loops. The usual approach is Yang-Mills regularization, in which a small kinetic term is added to the photon and then decoupled; in this approach, one finds that quantities in the $SU(\rnk)_\ktrue$ theory depend not on $\ktrue$ but $\ksh = \ktrue+\rnk \sign(\ktrue)$. The results of \cite{cpt-anyons,cpt-nonanyon} were presented in another regularization scheme (equivalent to dimensional regularization) in which there is no such shift. Therefore in importing results from there, we include this shift.
\item It may be that this duality does not hold for all numbers of flavours $N_f$. We will see that we require $N_f \le N$ for it to work out in our case, where $N$ is the number of colours in the bosonic theory.
\end{enumerate}

\subsection{The Spectra of Non-Relativistic CFTs}

The above theories both have the interesting property that a part of their spectrum is protected. (In the supersymmetric completion, this is the BPS portion of the spectrum.) Thinking of these theories as specifying the quantum mechanics of anyonic particles, if one adds a harmonic trap, one finds a class of states called \emph{linear states} whose energy depends linearly on the inverse Chern-Simons level. Alternatively, thinking of them as defining a CFT with an appropriate algebra of local operators, one finds that the dimensions of a certain class of operator depend on the level in the same way. These pictures are related by a state-operator map, which (unlike in the case of relativistic CFTs) simply maps operators in the original theory to states in the theory with a harmonic trap added. This part of the spectrum was analyzed in detail in \cite{cpt-anyons,cpt-nonanyon}.

One part of those papers was to understand what the operators of the CFT are. As a consequence of Gauss's laws, we are obliged to turn on the gauge field when turning on the matter fields -- indeed, in the example of \eqref{gauss-bose}, it is clear that excitations created by $\phi^\dagger$ carry magnetic charge.

The result is that the physical excitations of each system are best expressed in terms of dressed versions of the $\phi$ operators, carrying magnetic charge. But there is a simple way to achieve this if we sit on the plane. We simply dress $\phi$ with a Wilson line stretching out to infinity. This, being gauge invariant except under transformations at infinity, automatically satisfies Gauss's law. Hence we define
\begin{align}\label{gauge-invariant-field}
\Phi_{i}(\mathbf{x}) &= {\cal P} \exp\left(i\int_{\infty}^{\mathbf{x}} a + \tilde{a} \right) \phi_i(\mathbf{x})
\end{align}
and similarly $\Psi$, though for the latter we need to include an extra Wilson line for the extra gauge field, and adjust the charges to satisfy both Gauss laws. Path-ordered exponentials can also be seen in solutions to the Knizhnik-Zamolodchikov equations which arise naturally in anyon quantum mechanics; indeed, this is what imbues the particles with their anomalous statistics \cite{Lee1994}.

It is not immediately obvious that we can think of these as analogous to the \emph{local} operators of a relativistic CFT, due to the manifestly \emph{non-local} Wilson line, but in fact one can sensibly discuss things like the scaling dimension of such an operator, and arguments to that effect are given in \cite{cpt-nonanyon}. (One key reason for this is the difference in the state-operator map. In the relativistic case, the states live on a sphere, and this means we have nowhere to run the Wilson lines to. The non-relativistic harmonic trap has no such issue.) This means that these $\Phi$ fields are useful objects to discuss, though their extended nature means they have some significant subtleties (related to fusion rules, which constrain how they may be combined) to which we will return later.

Since we are interested in operators at a point, from here on we will understand operators to be evaluated at the origin, so $\Phi_i = \Phi_i(\mathbf{0})$ and so forth. Then, letting $\partial = \partial_z$ and $\bar{\partial} = \partial_{\bar{z}}$, the operators we are interested in are linear combination of terms like
\begin{equation*}
\cO_{i_1\cdots i_p} = \prod_{m=1}^p \partial^{l_m}\bar{\partial}^{l'_m} \Phi^\dagger_{i_m}
\end{equation*}
and similarly for $\Psi$ in the fermionic theory. Note that these transform in some representation $R$ of the global part of the gauge group as well as the $SU(N_f)$ flavour symmetry, but we have suppressed the indices of the former.

Note also that operators which are total derivatives are (in a sense made precise in terms of the conformal algebra in \cite{cpt-anyons,cpt-nonanyon}) \emph{descendants} of simpler operators. We ignore these in the following.

Now one can easily express the protected operators in terms of the $\Phi_i,\Psi_i$. We find that they are simply those with only $\partial$ derivatives, so $l'_m = 0$. In the bosonic theory, they look like this:
\begin{equation*}
\cO_{i_1\cdots i_p} = \prod_{m=1}^p \partial^{l_m} \Phi^\dagger_{i_m}
\end{equation*}

The dimension of such an operator is given by
\begin{equation*}
\Delta = p - (\cJ - ps)
\end{equation*}
where $\cJ$ is its angular momentum, appropriately regularized by subtraction of each individual particle's spin. One may think of this essentially as a measure of the binding energy of the composite operator. (This saturates a lower bound on the dimension of any operator with these eigenvalues of the number operator and angular momentum.)

Moreover, there is an elegant formula for this angular momentum. There is clearly some angular momentum associated to the derivatives $\partial^{l_m}$ in the above expression for $\cO$, so let us account for this by defining
\begin{equation}
\cJ - ps = - \sum_{m=1}^{p} l_m + \cJ_0 \fstp
\end{equation}
The more interesting thing is what remains. It is helpful to consider each factor of the gauge group separately. Suppose the gauge group is a product of the groups $G_I$ at levels $k_I$. Then take $\cO$ to be in a definite representation $R_I$ of each factor. Then
\begin{equation} \label{j0-general}
\cJ_0 = - \sum_I \frac{C^{(I)}_2(R_I) - p \, C^{(I)}_2(\text{fund})}{2\ksh_I}
\end{equation}
where $C_2(\cdot)$ is the quadratic Casimir of a gauge group representation, defined by
\begin{equation}
\sum_{\alpha}t^\alpha[R]t^\alpha[R] = C_2(R) \mathbf{1}
\end{equation}
for $t^\alpha$ generators in the appropriate representation, normalized by $\tr t^\alpha t^\beta = \delta^{\alpha\beta}$. We have written the solution in terms of the renormalized level $\ksh_I$ for each factor.

The spin $s$ of the individual particles is given by a related formula:
\begin{equation}
s = -\sum_I\frac{C^{(I)}_2(\text{fund})}{2\ksh_I}
- \begin{cases}
0 & \text{`bosonic' theory} \\ \frac{1}{2} & \text{`fermionic' theory}
\end{cases}
\end{equation}
For example, consider an Abelian theory with $\hat{k} = 1$. We see that bosons pick up a half-integer spin and fermions an integer spin, as expected.

From this point of view, it is clear that something fairly non-trivial must happen to match these dimensions across the duality. This is what we will describe in section \ref{sec-protected}.

\subsection{Two Dimensional CFTs and Bosonization}

The above formulae are on the one hand specializations of the results of \cite{cpt-nonanyon} as those hold for $\phi$ in arbitrary representations of the gauge group; although on the other hand, they are also slight generalizations of those since here we include extra gauge group factors. However, we note also that these formulae are clearly intimately related to the so-called \textit{minimal energies} of the affine Lie algebra representations associated to $R_I$ \cite{kac_1990}, or the conformal dimensions of the operators in WZW theory.

The key idea is one we have already encountered, in fact, in \eqref{gauge-invariant-field}. This, roughly speaking, takes the form of a \textit{decoupling transformation} which separates the Chern-Simons field theory from the matter fields. Similar transformations are used in the free fermion construction of Wess-Zumino-Witten theories in 2 dimensions \cite{naculich,cabra}. They also make direct contact with old ideas about the relationship of Chern-Simons theory on manifolds with boundary to conformal field theory \cite{wittenknot}.

Recall $a_z$ is flat away from $\phi$ insertions. Suppose there were no $\phi$ insertions. Then we could write $a_z = i h^{-1} (\partial_{z} h)$ for some $h$, and
\begin{equation*}
\partial_z (h \phi) = h\partial_z \phi + (\partial_z h) \phi = hD_z \phi
\end{equation*}
and hence one can write the action elegantly in terms of the fields $\tilde{\phi} = h \phi$ and $h$, with the Hamiltonian \eqref{hams} becoming simply $H\propto\int \mathrm{d}^2 x \ |\partial_z \tilde{\phi}|^2$. Moreover, these two fields are completely decoupled, except for Gauss's law, which would impose boundary conditions on $h$ at each $\tilde{\phi}$ insertion. The path integral, in this sector, reduces to a Wess-Zumino-Witten theory for $h$ on the boundary at infinity. Of course, this is a very trivial observation without any matter insertions!

As we add $\phi$ insertions, we start obtaining multiple boundary components, and we cannot typically expect a single-valued $h$. In general, this can result in a much more complicated path integral; indeed, even for Abelian anyon theories, not much is known about the dimension of generic operators. However, for the special case of chiral operators we obtain exactly the construction of 1+1 WZW models alluded to in the last section of \cite{wittenknot}, with a WZW field $h$ decoupling from some $\tilde{\phi}$ source as above. Accordingly, we can understand these 3d chiral operators in terms of WZW chiral primaries, and this underlies the nice formula \eqref{j0-general} for the corresponding operator dimension.

Of course, once we understand that there is a link between the chiral operators we are studying and operators in WZW models, we find ourselves in the world of level-rank dualities \cite{levelrank} and non-Abelian bosonization \cite{Witten-nonab} -- thus it is perhaps no surprise that we can understand the duality of this sector of the 3d theory in terms of level-rank dualities in 2d. The goal of this paper, then, is to elucidate precisely what this duality looks like in terms of the 3d fields.

Finally, as mentioned in the introduction, we note that in \cite{cpt-nawzw} a precise relationship was established between the chiral sector of a deformation of this CFT and a WZW model. Specifically, a chemical potential was added, and the moduli space of $v$ vortices in a harmonic trap was studied. The dynamics within this moduli space, as $v\to\infty$, was shown to be precisely that of the chiral Wess-Zumino-Witten theory by matching both the algebra and the partition function of the former to the latter. This is clearly similar to what we have seen here: the chiral sector of the theory matches a chiral WZW model.


\clearpage
\section{A Recipe for Matching Protected Operators}\label{sec-protected}

We will describe the matching for protected operators in three stages. Firstly, we will observe that the single-particle operators match. Secondly, we will explain how ``small'' multi-particle states, namely those which avoid triggering the fusion rules, also match. Finally, we will explore how arbitrary multi-particle states match.

Before we embark on this path, it is worth emphasizing that we are going to provide a matching not only between gauge singlet operators, but actually between gauge-dependent operators. This obviously is not a bijection between operators (consider the single-particle operators in e.g. a $U(N) \leftrightarrow U(1)$ duality), but it is a bijection between their gauge orbits. We offer two perspectives upon this:

\begin{itemize}
\item The first the point of view is that of Chern-Simons theory. Let us consider putting some charges into a Chern-Simons field theory, and asking that the state transforms under some particular representation of the gauge group at infinity. We can do this by inserting Wilson lines at the appropriate points. Then, as famously discussed in \cite{wittenknot}, one can analyze the Hilbert space of possible Chern-Simons states which can exist given these charges. These spaces are the objects we match.
\item The second is completely equivalent, but phrased in condensed matter language. By coupling matter to a Chern-Simons gauge theory, we are really looking at theories of \textit{anyons}. With each physically distinct representation of a group $SU(N)$, we associate a single \textit{species} of anyon. Thus in this language, our duality is an implementation of the level-rank duality of anyon species.
\end{itemize}
We emphasize that it is known that, for instance, the \textit{fusion rules} dictating how anyon species can fuse into new compound anyons are preserved across level-rank duality \cite{yellow}. This is essentially the same thing as the statement that the Hilbert spaces of Chern-Simons states agree across level-rank duality. Our goal in this paper is to give descriptions of the operators which create such states, demonstrating that their dimensions and angular momenta agree across the duality.

Note that there are two ways to handle what happens at infinity. One option is to compactify and think of inserting a single Wilson line there too. The other is to dress treat the theory as having a boundary, and add a WZW theory containing operator insertions at that boundary. In either case, the fusion rules also constrain what representation the insertion (and therefore the bulk state) may transform in. The duality will match partition functions up to the level-rank duality of this global representation.

With that all said, let us begin by considering the situation with single-particle operators. The only possible matching is
\begin{equation}
\cO_i = \partial^l \Phi_i^\dagger \qquad \longleftrightarrow \qquad \tilde{\cO}_i = \partial^l \Psi_i^\dagger
\end{equation}
and this indeed works out very neatly. The dimensions of both operators (which are free) match trivially. The more pleasing thing is that their angular momenta are equal. For the case of $n=1$, this is realized simply by the identity
\begin{equation}
\cJ_{\Phi^\dagger} = - \frac{N}{2 (k+N)} = - \frac{k}{2\times- (N+k)} - \frac{1}{2} = \cJ_{\Psi^\dagger} \fstp
\end{equation}

But what about multi-particle states, where the anomalous (binding-energy-like) angular momenta kick in?

\subsection{``Small'' Multi-Particle Operators}

Suppose that you have a representation $R$ of some unitary group $U(N)$. Then we can describe it efficiently using an unreduced Young diagram with $\lambda_i$ boxes in the $i$th row. Here, ``unreduced'' means we allow full columns of $N$ boxes in order to keep track of the total $U(1)$ charge too.

As mentioned above, it is simplest and perfectly general to focus on the $n=1$ case, since here the $U(1)$ level and the renormalized $SU(N)$ level agree.

The corresponding anomalous angular momentum associated with this diagram can be evaluated by using the expression for the quadratic Casimir in terms of the inner product of weights $\left<\lambda, \lambda + 2\rho\right>$. (Here, $\rho$ is the sum of fundamental dominant weights.) One obtains
\begin{equation}\label{j0-lambda}
\cJ_0 = - \frac{C_2(R) - p \, C_2(\text{fund})}{2\ksh} = - \frac{\sum_i  \left[  \frac1 2 \lambda_i (\lambda_i - 1) - (i-1) \lambda_i \right] }{\ksh}
\end{equation}
where we have written this last term in a very particular way, for reasons that will now become clear.

The protected operators we have discussed can be constructed by taking an operator built from the right number of bosons and derivatives, and contracting its gauge indices $a_m$ with a tensor $M$ transforming in the representation $R$:
\begin{equation}
\cO_{i_1 \cdots i_p} = M_{a_1 \cdots a_p} \partial^{l_1} \Phi_{i_1 }^{\dagger a_1} \cdots  \partial^{l_p} \Phi_{i_p }^{\dagger a_p}
\end{equation}
We seek a dual for this operator, constructed from fermions, which transforms in the same $SU(N_f)$ representation. Let us consider operators of the form
\begin{equation} \label{simpledual}
\tilde{\cO}_{i_1 \cdots i_p} = \tilde{M}_{a_1 \cdots a_p} \partial^{l_1} \Psi_{i_1 }^{\dagger a_1} \cdots  \partial^{l_p} \Psi_{i_p }^{\dagger a_p}
\end{equation}
Replacing bosons with fermions means that everything which was symmetrized is now anti-symmetrized and vice versa. So if we want the same $SU(N_f)$ representation, we must put $\tilde{M}$ in a representation $\tilde{R}$ described by a partition $\lambda^T$:
\begin{equation}
\Yvcentermath1 \lambda = \yng(7,5,2,2,1) \qquad \longleftrightarrow \qquad \lambda^T = \yng(5,4,2,2,2,1,1)
\end{equation}
This, of course, is not a surprise: level-rank duality generally relates Young diagrams to their transpositions, up to the important issues to be discussed in section \ref{singlet-sec}.

What is the anomalous angular momentum $\cJ_0$ of this state? It is easy to work out by looking at \eqref{j0-lambda} the right way. The first term in the sum counts the number of pairs of boxes lying in the same row of the Young diagram. The second similarly counts the number of boxes lying in the same column. But upon transposition, these two terms are simply interchanged.

Moreover, recall that under the $n=1$ duality, $\hat{k} \to -\hat{N}$, but $\hat{k} = \hat{N} = k + N$ when both quantities are positive. Hence in fact, $\cJ_0 \to \cJ_0$ is invariant for these theories!

This is the first clear manifestation of the multi-particle bosonization duality: the anomalous angular momentum for an bosonic operator and its naive dual \eqref{simpledual} are identical. It follows that their dimensions $\Delta$ and angular momenta $\cJ$ are also equal. This result essentially reproduces old observations about the behaviour of Wilson line observables under level-rank duality \cite{levelrank}.

We see that given a $\lambda$ describing a valid representation of $SU(N)$ then the transpose $\lambda^T$ is a valid representation of $SU(k)$ only if it has at most $k$ columns. But this (neglecting $SU(N)$ singlets for the moment) is precisely the requirement that the representation is an \textit{integrable} representation of $SU(N)_k$.

This notion will be important for us. In general, when we bring operators transforming under a gauge group with a finite level, rather than transforming in the tensor product, the composite operator transforms only in the representations specified by the fusion rules of that theory \cite{yellow}, as mentioned in the introduction. In particular, $SU(k)$ representations whose reduced Young diagrams exceed $k$ columns are always projected out. A nice characterization of the fusion rules is that when one brings together a fundamental weight (a partition which is a column of boxes) with some other integrable partition, one simply takes the normal tensor product but then sets to zero any partition whose Young diagram is not integrable.

This is the key subtlety associated with the procedure for attaching Wilson lines we postulated above. Thinking of the physical states as being given by Chern-Simons theory defined around fixed point charges, the Hilbert space of gauge configurations is only non-empty when we obey the fusion rules, as in the story told in \cite{wittenknot}.

In summary, if we start with a $\lambda$ describing a valid representation of $SU(N)_k$, with no singlet factors, then the transpose $\lambda^T$ is always a valid representation of $SU(k)_N$. But we do not have a procedure which works if we add many singlets to the operator.

\subsection{Adding Singlets}\label{singlet-sec}
The above is clearly is only a part of the story. In general, there is no operator of the form \eqref{simpledual}, because if we include several $SU(N)$ singlets in $\lambda$, then there may be too many antisymmetrizations in $\lambda^T$ for it to be a valid representation of $SU(k)$. But in order for the formulae for angular momenta and dimensions to be equal, we must include these singlets to get the correct $U(1)$ contributions.

From a more mathematical point of view, there is no bijection between representations of affine special unitary Lie algebras. The usual level-rank duality picture only specifies a bijection between representations of $SU(N)_k$ and $SU(k)_N$ modulo \textit{outer automorphisms}. The outer automorphism groups are respectively $\mathbb{Z}_N$ and $\mathbb{Z}_k$, and are associated with the center of each original group.

The action of $\mathbb{Z}_k$ upon reduced representations of $SU(k)$ at level $N$ is easy to describe: the generator $\sigma$ adds a row of $N$ boxes to the top of the diagram and then reduces it. (It is easy to verify $\sigma^k = 1$.) But from the point of view of the bosonic side of the duality, clearly this looks much like adding a column of $N$ boxes -- that is, adding a singlet -- before then performing a rather mysterious operation, namely removing complete rows of $k$ boxes.\footnote{Note that the matching between singlets and maximally symmetric representations corresponds to the matching between baryons and monopoles discussed in e.g. \cite{ofer}.}

This all suggests that the $U(1)$ part of the gauge group must mix in some non-trivial way with some other source of angular momentum. This other source can only be derivatives.

To see this, it makes sense to focus firstly on Abelian theories, where things are simple and we only have these $U(1)$ factors to worry about. Thus we will set $N=k=N_f$ = 1, but keep $n$ general. Consider the lowest-dimension protected operator in the bosonic theory:
\begin{equation}\label{abab-bos-grnd}
\cO = \underbrace{\Phi^\dagger \Phi^\dagger \cdots \Phi^\dagger}_p \qquad \text{ in }\qquad U(1)_{n+1}
\end{equation}
This should presumably match the lowest-dimension fermionic operator,
\begin{equation}\label{abab-frm-grnd}
\tilde{\cO} = \underbrace{\Psi^\dagger \partial \Psi^\dagger \cdots \partial^{p-1}\Psi^{\dagger}}_p \qquad \text{ in }\qquad U(1)_{-1} \times U(1)_{n} \equiv U(1)_{-1-1/n}
\end{equation}
where we have integrated out the extra $U(1)_n$ for expedience, this being equivalent to doing a more careful calculation of dimensions.
Indeed, this pans out very neatly:
\begin{align*}
\Delta_{\cO} &= p + \frac{p(p-1)}{2(n+1)} \\
\Delta_{\tilde{\cO}} &= p + \frac{1}{2}p(p-1) - \frac{np(p-1)}{2(n+1)} = \Delta_{\cO}
\end{align*}

This makes sense: the statistical parameters of these theories are $\theta = \pi/(n+1)$ and $\tilde{\theta} = -n\pi/(n+1)$ respectively, and they are related by $\theta = \tilde{\theta} + \pi$, or the distinction between bosons and fermions.

Focussing on the $n=0$ case for simplicity, this is the standard flux attachment procedure for turning bosons into free fermions. This can also be expressed as the identity
\begin{equation}\label{abelian-flux-index}
q^{x^2 \partial_x^2/2} \prod_{m=0}^\infty \frac{1}{1-x q^m} = \prod_{m=0}^\infty (1+x q^m)
\end{equation}
which equates chiral indices for the two theories. (These are the partition functions of the chiral operators in the two theories, $Z=\operatorname{Tr}_{\rm chiral} [x^{\cal N} q^{\Delta-{\cal N}}]$, where ${\cal N}$ is the number operator. Each term in the product deals with $m$-derivative single particle operators, whilst the prefactor on the left attaches flux to the bosons.) Writing $(x;q)_n=\prod_{l=0}^{n-1} (1-xq^k)$ for the $q$-Pochhammer symbol, we see that this identity combines two common special cases of the $q$-binomial theorem:
\begin{equation}
\frac{1}{(x;q)_\infty} = \sum_{n=0}^\infty \frac{x^n}{(q;q)_n} \qquad \text{and} \qquad
\sum_{n=0}^\infty \frac{q^{n(n-1)/2}}{(q;q)_n}x^n = (-x;q)_{\infty}
\end{equation}
The usual combinatorial interpretation of these results is precisely that partitions of an integer $n$ into integers and \textit{distinct} integers are differentiated precisely by the inclusion of a triangular partition -- in physics language, bosons and fermions are related by the inclusion of $\partial^0,\partial^1,\ldots,\partial^{n-1}$ derivatives included in the chiral fermionic ground state.

Here, we see clearly that indeed there is a mixing between explicit and anomalous angular momentum across the duality. This is the key idea we need to identify the matching between operators.

However, things are more subtle when we have a non-trivial $SU(N)$ part to the gauge group. When the group sits at level $k$, we are forbidden from constructing operators transforming as (reduced) Young diagrams with more than $k$ columns. From the point of view of our unreduced Young diagrams, we may have more than $k$ columns, but only by adding singlets (columns of $N$ boxes) to the left of the diagram.

From the point of view of the dual $SU(k)_N$, of course, added singlets appear as extra maximal-length rows of $N$ boxes at the top of the reduced Young diagram. Generically the diagram then needs reducing to be a $SU(k)$ state. This is how we explore the orbit of the outer automorphism group under the duality.

This means that, by tracking the total $U(1)$ charge of the states transforming in these representations, we actually obtain the outer automorphisms much more naturally than in the above presentation. One simple way to understand why this is from the embedding
\begin{equation}
u(1)_{Nk} \oplus su(k)_{N} \oplus su(N)_{k} \subset u(Nk)_{1}
\end{equation}
commonly discussed in the context of level-rank duality.
Consider the free fermion representation of the right-hand side; in fact, simply take a quantum mechanics of free fermions $\chi_{a}^i$ transforming in $SU(k) \times SU(N)$. The resulting states decompose into representations of the left-hand side. One easily sees that things in the representation $\lambda$ of $SU(N)$ only appear in representations $SU(k)$ which are $\lambda^T$ or elements $\sigma^r(\lambda^T)$ of its outer automorphism orbit, precisely because of the potential presence of singlets. But clearly the $U(1)$ charge precisely counts the number of singlets, and hence disambiguates the level-rank map. The $U(1)$ and $SU(k)$ representation together specify the $SU(N)$ representation uniquely; and similarly with $k \leftrightarrow N$.

We need to check that this allows us to construct some sort of bijection which preserves the angular momenta and dimensions.

\subsubsection*{Constructing a Matching Operator}

Suppose you are given a bosonic operator consisting of $qk+r$ singlets and a part transforming under a reduced $SU(N)$ representation $\mu$ which is integrable (i.e. with at most $k$ columns). There are $q$ blocks which are $N\times k$ rectangles, which upon transposition remain gauge singlets. Put these to the left of the new diagram. The remainder of the transposed diagram -- $r$ rows of $N$ blocks atop $\mu^T$ -- may be invalid if it has more than $k$ rows. If it does, take the portion of the diagram below the $k$th row and place it instead at the top right of the diagram. This is necessarily a valid $U(k)$ diagram, and also has at most $N$ columns which are not singlets.

\begin{figure}
\begin{subfigure}{\textwidth}
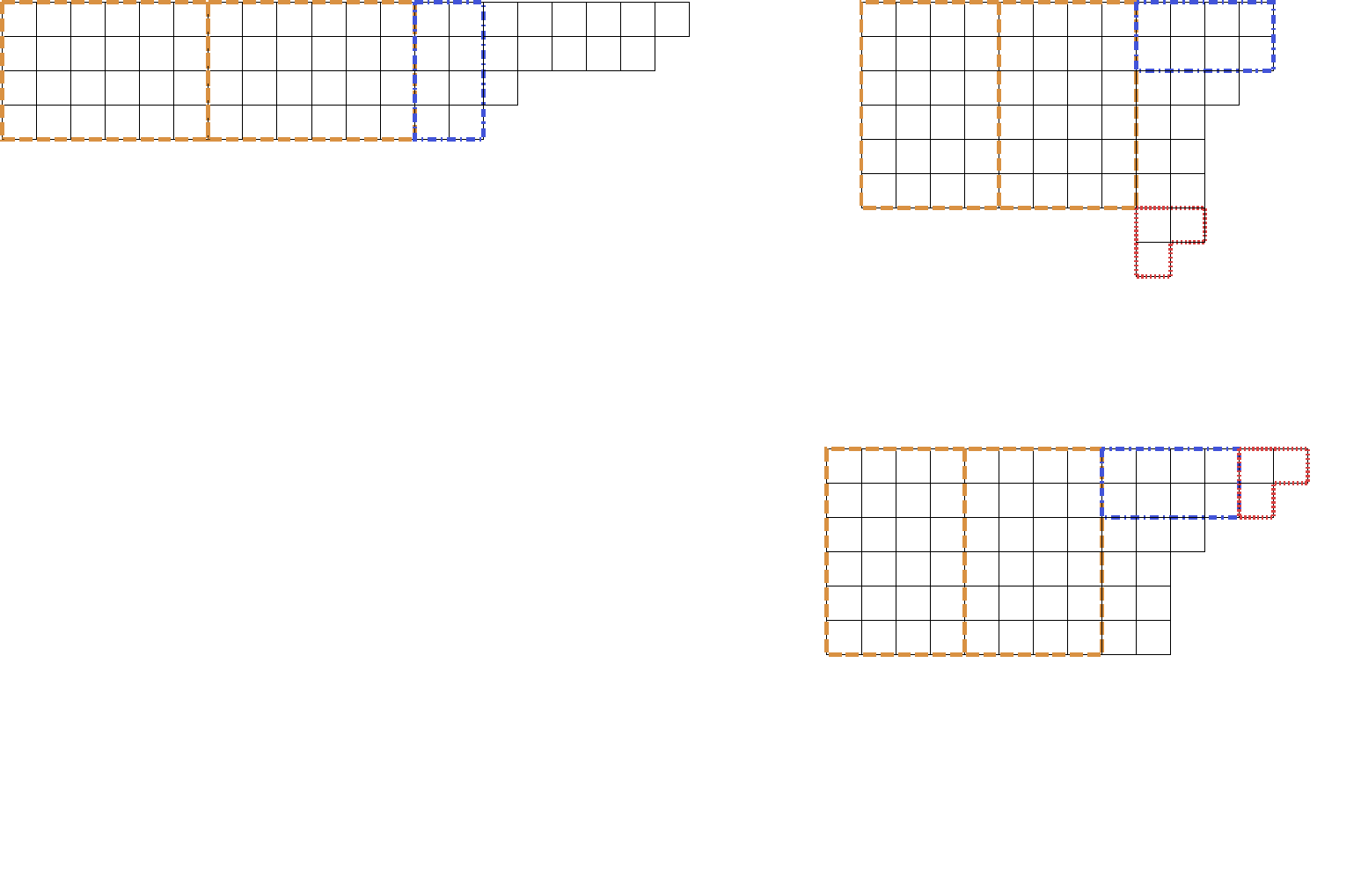
\caption{The $U(4) \to U(6)$ description}
\end{subfigure}
\begin{subfigure}{\textwidth}
\vspace{2em}
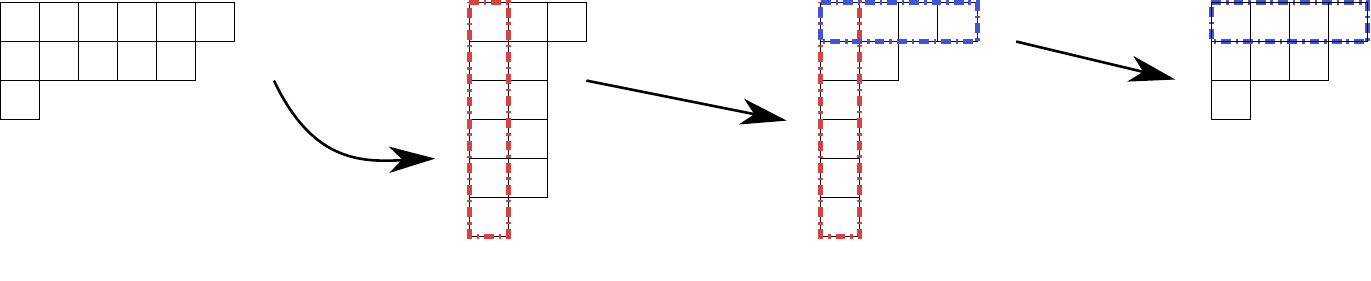
\caption{The corresponding $SU(4) \to SU(6)$ description}
\end{subfigure}
\caption{\label{figmap} Transposing from $U(4) \supset SU(4)_6$ to $U(6) \supset SU(6)_4$}
\end{figure}

An example of this process is depicted in Figure \ref{figmap}. Note also that the final $SU(k)$ representation is related to the transpose of the original $SU(N)$ representation by the outer automorphism $\sigma^r$.

We must now dress the operator with the appropriate derivatives so that a valid fermionic object may be constructed transforming in this way. The recipe is simple enough: divide the completed diagram into sets of $N$ columns labelled by $I=1,2,\ldots$. Then every operator sitting in the $I$th region must have $I-1$ derivatives added to it.

To reverse this process, and go from fermions to bosons, we simply remove $I-1$ derivatives instead of adding them.

For instance, if one starts in a theory with $N_f = N = 3$ and $k=2$ and a bosonic operator like
\begin{equation}
\cO = \left(\Phi^{\dagger[a_1}_{[i_1} \Phi^{\dagger a_2}_{i_2} \Phi^{\dagger a_3]}_{i_3]}\right)^4
\end{equation}
then the fermionic dual is 
\begin{equation}
\cO = \left(\Psi^{\dagger(a_1}_{[i_1} \Psi^{\dagger a_2}_{i_2} \Psi^{\dagger a_3)}_{i_3]}\right)^2 \left(\partial\Psi^{\dagger(b_1}_{[j_1} \partial\Psi^{\dagger b_2}_{j_2} \partial\Psi^{\dagger b_3)}_{j_3]}\right)^2
\end{equation}
where each square contains an $SU(2)$ antisymmetrization to make it a singlet. Here, the first squared bracket is labelled $I=1$ and the second square is labelled $I=2$, hence the added derivatives.

There are two questions to answer. Firstly, do the quantum numbers still match after this transformation? Secondly, does this really give a bijection between non-vanishing operators? We will begin by proving that the answer to the first question is yes. The second question proves more subtle, and will require some more work to understand.

\subsubsection*{Proof that Quantum Numbers Match}
We can think of the process depicted in Figure \ref{figmap} as simply transposing and then moving each $N\times k$ block alongside the previous one instead of lying on top of it. This is emphasized by the labelling of these blocks as $I=1,2,\ldots$.

We only ever move blocks of cells that are in columns containing a multiple of $k$ other cells, and we only ever place them alongside rows containing a multiple of $N$ other cells. This guarantees that each cell we moves shifts $\cJ_0 \sim (\text{column pairs} - \text{row pairs})/(N+k)$ by an integer. Moreover, if we count pairs by summing for each cell the number of cells above it or to its left, that integer is easily seen to be $(I-1)$ for each cell in the $I$th region which we move. \hfill\qedsymbol

\subsection{A Corresponding Index Identity}\label{indexidentity-sec}

Constructing a corresponding index which matches for all these gauge-non-invariant objects is now a little more subtle. It is illuminating (though not strictly necessary) to proceed by first starting with the index for free scalars or fermions, and then describing the effect of flux attachment on the index.

Let us focus on the scalar theory, and for simplicity of the notation let us restrict to the case of $N_f = 1$ (though a straightforward generalization is available). Then with a non-Abelian $U(N)$ symmetry, ungauged to begin with, we have
\begin{equation}
Z_{\rm free} = \prod_{a=1}^{N}\prod_{m=0}^\infty \frac{1}{1 - \omega_a q^m}
\end{equation}
where $\omega_a$ are fugacities for the $U(N)$ Cartan elements, so $a=1,\ldots,N$. Expanding this out, each power of $\omega_a q^m$ corresponds to a $\partial^m \Phi^{\dagger a}$.

Now the effects of the flux attachment are best understood in a basis of irreducible representations of the $U(N)$. This means we want to decompose $Z_{\rm free}$ into a sum over Schur polynomials,
\begin{equation}\label{schur-decomp}
Z_{\rm free} = \sum_{\lambda} S_{\lambda}(\omega) f_{\lambda}(q)
\end{equation}
where we sum over all partitions $\lambda$ with at most $N$ parts, and $S_\lambda$ are the Schur polynomials
\begin{equation}
S_\lambda(\omega) = \sum_{\sigma \in S_N} \sigma \left\{ \ \omega_1^{\lambda_1} \omega_2^{\lambda_2} \cdots \omega_N^{\lambda_N} \ \prod_{a>b} \frac{1}{\left( 1-\frac{x_i}{x_j} \right)} \  \right\} \fstp
\end{equation}
These precisely correspond to irreducible representations of $U(N)$. The infinite series $f_\lambda(q)$ describes the dimensions of all the chiral operators in this representation.

There are two key effects of the flux attachment which we now include. Firstly, as we have established, there is an anomalous dimension $J_0[\lambda]$ associated with the gauging; in $U(N)_{k,k+N}$ this was
\begin{equation*}
J_0[\lambda] = - \frac{\sum_i  \left[  \frac1 2 \lambda_i (\lambda_i - 1) - (i-1) \lambda_i \right] }{k+N}
\end{equation*}
for instance. This shifts $S_\lambda \to q^{-J_0[\lambda]} S_\lambda$ in \eqref{schur-decomp}. Note that we implement this even for non-integrable representations.

Secondly, the fusion rules must be included. It might be rather surprising that this can be implemented effectively starting from the free spectrum given that most of the representations in \eqref{schur-decomp} are not integrable and must drop out of the spectrum. In fact, it transpires that this works out very elegantly. All we need a recipe for taking an arbitrary $U(N)$ representation and spitting out the corresponding level $k$ representation.

It turns out that this is a standard procedure in the representation theory of affine Lie algebras. (See section 16.2.2  of \cite{yellow}.) The mathematical picture is that the weight space for the affine Lie algebra is decomposed into \textit{affine Weyl chambers}. One takes all non-integrable weights, and reflects them using Weyl reflections back into the affine fundamental chamber. This results in some new representation ${\cal R}^{(k)}\lambda$ say, at the cost of $r$ reflections. One must then include ${\cal R}^{(k)}\lambda$ with the sign $\epsilon_\lambda = (-1)^r$ in the fusion rules. Weights lying on the boundary of the fundamental chamber are left invariant by a reflection and so `contribute with both signs'; we take $\epsilon_\lambda = 0$. Even though some representations appear with minus signs, they always cancel out other terms in the underlying tensor product, leaving simply a subset of the full tensor product.\footnote{A simple Young diagram manipulation version of this is described in section 16.2.4 of \cite{yellow}, and a Mathematica version is available at \url{http://blog.suchideas.com/2mBUW}. Note that normally this is discussed for $SU(N)$, neglecting singlets; one finds that simply re-inserting the appropriate number of singlets so as not to change the total $U(1)$ charge extends the algorithm to $U(N)$.}

Putting this all together, we have the recipe
\begin{equation}
S_\lambda(\omega) \quad \longrightarrow \quad \epsilon_\lambda \ q^{-J_0[\lambda]} \ S_{{\cal R}^{(k)}\lambda}(\omega)
\end{equation}
which can of course be implemented by integrating \eqref{schur-decomp} against an appropriate kernel in a generalization of the usual approach for picking out gauge singlet operators, since $\left<S_\lambda,S_\mu\right> = \delta_{\lambda\mu}$ for an appropriately chosen inner product.

Indeed, we can write the partition function for chiral states in the integrable representation $\mu$ of $U(N)$ as 
\begin{equation}
Z_\mu = \frac{1}{N!} \left( \prod_{i=1}^{N} \frac{1}{2\pi i} \oint \frac{\rmd \omega_a}{\omega_a} \right) \prod_{a<b}\left(1-\frac{\omega_a}{\omega_b}\right) \ \sum_{\lambda : {\cal R}^{(k)} \lambda = \mu }^{} \epsilon_\lambda q^{-J_0[\lambda]} S_\lambda(\omega^{-1})  Z_{\rm free}
\end{equation}
where $\omega^{-1} = (\omega_1^{-1}, \ldots, \omega_N^{-1})$.

One can do exactly the same construction for fermions in $U(k)_{-N,-N-k}$ to obtain a partition function $\tilde{Z}_{\tilde{\mu}}$ for each integrable representation $\tilde{\mu}$ of $U(k)$ at level $N$. Then we find that indeed
\begin{equation}\label{partitionequality}
Z_\mu = \tilde{Z}_{\tilde{\mu}}
\end{equation}
where $\tilde{\mu}$ is related to $\mu$ in precisely the way described in section \ref{singlet-sec}. This is presumably very closely related to the usual WZW level-rank dualities.

The arguments above form much of the proof of this identity; we will not provide a detailed proof here, instead electing to illustrate it with some simple examples.

\clearpage
\section{Examples and Analysis}\label{analysis-sec}

It is productive to take this opportunity to explore the details of this matching in the simplest possible settings. We will first look at the single flavour indices, and then continue to examine the operator matching in detail for larger $N_f$.

\subsection{Index Examples and A General Lesson}\label{index-examples}

We will start by illustrating the rules about $N_f = 1$ indices presented in section \ref{indexidentity-sec} in a very simple non-Abelian setting.

\subsubsection*{Example: $N=2,k=1$}

The simplest possible non-Abelian example is $U(2)_{1,3}$ with a single boson. Here, we find that the free boson index looks like
\begin{equation*}
\Yboxdim7pt
Z_{\rm free} = 1 + \frac{1}{1-q} Z_{\rm free}'
\end{equation*}
where we separate the singlet then factorize out total derivatives to leave only primary operators
\begin{align*}
Z_{\rm free}' &=
S_{\yng(1)}
+ \frac{1}{1-q^2} S_{\yng(2)}
+ \frac{q}{1-q^2} S_{\yng(1,1)}
+ \frac{1}{(1-q^2)(1-q^3)} S_{\yng(3)} \\& \quad
+ \frac{q}{(1-q)(1-q^3)} S_{\yng(2,1)}
+ \frac{1}{(1-q^2)(1-q^3)(1-q^4)} S_{\yng(4)} \\& \quad
+ \frac{q}{(1-q)(1-q^2)(1-q^4)} S_{\yng(3,1)}
+ \frac{q^2}{(1-q^2)^2(1-q^3)} S_{\yng(2,2)} + \cdots
\end{align*}
where we have listed all operators with up to 4 $\Phi$ insertions. Note that the denominator of each term can be understood in terms of products of $(1-q^h)$ where $h$ ranges over all hook lengths, and we have factored out the trivial $1/(1-q)$ present in all diagrams. The numerators count the number of antisymmetrizations required (i.e. the minimum number of derivatives needed to prevent the corresponding operator vanishing).

Now for $k=1$, many of these operators are affected by the fusion rules. One finds that
\begin{align*}
\Yvcentermath1\yng(2)  \qquad&\longrightarrow\qquad 0 \\
\Yvcentermath1\yng(3)  \qquad&\longrightarrow\qquad - \ \Yvcentermath1\yng(1) \\
\Yvcentermath1\yng(4)  \qquad&\longrightarrow\qquad - \ \Yvcentermath1\yng(2,2) \\
\Yvcentermath1\yng(3,1)  \qquad&\longrightarrow\qquad 0
\end{align*}
where we have not yet included the $q^{-J_0}$ terms.

Hence the physical partition function, as a sum over all integrable partitions, but dropping the vacuum state and the total derivatives as we did for $Z_{\rm free}'$, is as follows:
\begin{align*}
Z' &=
S_{\yng(1)}
+ \frac{q \times q^{-1/3}}{1-q^2} S_{\yng(1,1)}
+ \left(\frac{q \times q^0}{(1-q)(1-q^3)} - \frac{1\times q^{1}}{(1-q^2)(1-q^3)} \right) S_{\yng(2,1)} \\& \quad
+ \left(\frac{q^2 \times q^0}{(1-q^2)^2(1-q^3)} - \frac{1 \times q^2}{(1-q^2)(1-q^3)(1-q^4)} \right) S_{\yng(2,2)} + \cdots \\
&= 
S_{\yng(1)}
+ \frac{q^{2/3}}{1-q^2} S_{\yng(1,1)}
+ \frac{q^2}{(1-q^2)(1-q^3)} S_{\yng(2,1)} + \frac{q^4}{(1-q^2)(1-q^3)(1-q^4)} S_{\yng(2,2)} + \cdots
\end{align*}

But now consider the dual theory of $U(1)_{-2}$ fermions. We find that
\begin{align*}
\tilde{Z}' &=
S_{\yng(1)}
+ \frac{q \times q^{-1/3}}{1-q^2} S_{\yng(2)}
+ \frac{q^3 \times q^{-1}}{(1-q^2)(1-q^3)} S_{\yng(3)} + \frac{q^6 \times q^{-2}}{(1-q^2)(1-q^3)(1-q^4)} S_{\yng(4)} + \cdots
\end{align*}
and clearly we indeed see that $Z' \equiv \tilde{Z}'$ in the sense of the identification \eqref{partitionequality}.

\subsubsection*{Example: $N=2,k=2$}

By contrast, if we go to level 2, then the relevant fusion rules become
\begin{align*}
\Yvcentermath1\yng(3)  \qquad&\longrightarrow\qquad 0 \\
\Yvcentermath1\yng(4)  \qquad&\longrightarrow\qquad - \ \Yvcentermath1\yng(3,1)
\end{align*}
which leads to
\begin{align*}
Z' &=
S_{\yng(1)}
+ \frac{q^{1/4}}{1-q^2} S_{\yng(2)}
+ \frac{q^{3/4}}{1-q^2} S_{\yng(1,1)}
+ \frac{q}{(1-q)(1-q^3)} S_{\yng(2,1)} \\& \quad
+ \frac{q^{5/2}}{(1-q)(1-q^3)(1-q^4)} S_{\yng(3,1)}
+ \frac{q^2}{(1-q^2)^2(1-q^3)} S_{\yng(2,2)} + \cdots \equiv \tilde{Z}'
\end{align*}
giving an elegant $SU(2)_2 \leftrightarrow SU(2)_2$ matching under which the second and third terms interchange but all other displayed terms are left invariant.

\subsubsection*{A General Lesson About Fusion Rules}

There is some useful information about how fusion rules are implemented buried in the above formulae.

Let us first look at $U(2)$ with $k=1$ and consider $p=3$ particle operators. The lowest dimension operator should be the operator with the fewest derivatives corresponding to the only possible $U(2)$ representation $\Yboxdim6pt \Yvcentermath1 \yng(2,1)$ which for $N_f=1$ is naively
\begin{equation*}
\mathcal{O} = \Phi^{\dagger [a} \partial \Phi^{\dagger b]} \Phi^{\dagger c}
\end{equation*}
with dimension $\Delta = 4$.
However, using our algorithm, the dual of this would be constructed as
\begin{equation*}
\tilde{\mathcal{O}} = \Psi^{\dagger} \partial \Psi^{\dagger} \partial\Psi^{\dagger}
\end{equation*}
which clearly vanishes. Thus either the correspondence is failing here, or these are not the correct operators to be looking at. Fortunately, the resolution is the less drastic second option.

Let us look at our index for this theory and try to get some insight. We see that the coefficient of $S_{\yng(2,1)}$ is shifted by the implementation of the fusion rules as
\begin{align}
\frac{q}{(1-q)(1-q^3)} &\longrightarrow \frac{q}{(1-q)(1-q^3)} - \frac{q}{(1-q^2)(1-q^3)}  \nonumber\\ &\ = \frac{q^2}{(1-q^2)(1-q^3)}
\end{align}
In particular, note that the first term in the series expansion is $q^2$ -- hence the lowest dimension operator in fact has \textit{two} derivatives, as it must do to match the fermionic side. Moreover, the whole rest of the spectrum is shifted significantly. The interpretation of the shift from the $q/(1-q) \to q^2/(1-q^2)$ suggests that a requirement to have ``at least one derivative, and possibly more'' is being replaced with a requirement to have ``at least one pair of derivatives, and possibly more pairs''. Why is this?

Consider the symmetrized operator $\mathrm{Sym}[\Phi^{\dagger a}(z_1)\Phi^{\dagger b}(z_2)\Phi^{\dagger c}(z_3)]$. Then the fusion rules insist that the only non-vanishing components of this as $z_1,z_2,z_3 \to z$ transform as $\Yboxdim6pt \Yvcentermath1 \yng(2,1)$ only. This is what we implemented above. Suppose instead we start bringing $z_1$ and $z_2$ together first, so that the wavefunction becomes proportional to $\epsilon_{ab}$; then for symmetry reasons, there must be some relative angular momentum, making the leading term proportional to $(z_1-z_2) \epsilon_{ab}$. This corresponds to constructing the operator $\Phi^{\dagger [a} \partial\Phi^{\dagger b]}$. But recall there is a third operator $\Phi^\dagger(z_3)$, symmetrized with the $z_1,z_2$ terms. If $(z_1-z_2)\epsilon_{ab}$ is to be the \textit{leading} term as we bring $z_1$ and $z_2$ together, then there must be some more angular momentum. Put another way, the operator written above appears to be order $O((z_1-z_3)^0)$ and so by symmetry it would also be  $O((z_1-z_2)^0)$.

Therefore, the first acceptable option can actually be written as
\begin{equation*}
\mathcal{O} = \Phi^{\dagger [a} \partial \Phi^{\dagger b]} \partial \Phi^{\dagger c} 
\end{equation*}
with the dual
\begin{equation*}
\tilde{\mathcal{O}} = \Psi^{\dagger} \partial \Psi^{\dagger} \partial^2 \Psi^{\dagger}
\end{equation*}
and the duality works! Both operators have dimension $\Delta = \tilde{\Delta} = 5$.

Moreover, the shifted excitation spectrum corresponds to the fact that any time we have an extra derivative on the $\Phi^{\dagger b}$ term, we must also add one to $\Phi^{\dagger c}$ -- such excitations indeed come in pairs. This matches with the same requirement on the fermionic side for the second two fermions.

This generalizes in a natural if sometimes subtle way; we must always include enough derivatives to guarantee that \textit{any subset of operators} may be brought together safely without violating the fusion rules. This means that if you take $q\le p$ operators, where you must take operators with fewer derivatives attached first, then it must be possible to choose those $q$ operators to obey the fusion rules. In the example above, consisting of the operators $\Phi^{\dagger a} \partial \Phi^{\dagger b} \Phi^{\dagger c}$, we must first pick the $\Phi^{\dagger a} \Phi^{\dagger c}$ terms, but these necessarily transform in the $a,c$ symmetric representation, violating the $k=1$ fusion rules.

\subsection{Simple Examples with More Flavours}

It turns out that there is something to say even about Abelian-Abelian dualities, so let us begin there, before continuing as before and looking at theories where instead one of the two factors is Abelian.

\subsubsection*{Two Particle Examples}

Consider $U(1)$ with $k=1$. Then for one flavour, the matching between \eqref{abab-bos-grnd} and \eqref{abab-frm-grnd} very straightforwardly generalizes to a complete matching of the whole spectrum, as the index identity \eqref{abelian-flux-index} demonstrates. But for $N_f = 2$ something immediately goes wrong. Consider the simple question: what is the bosonic dual of the fermionic operator $\tilde{\cO} = \Psi^{\dagger}_{[i} \Psi^{\dagger}_{j]}$? It is easy to verify that this (sticking to the $n=1$ duality $U(1)_2 \leftrightarrow U(1)_{-2}$) has dimension $\tilde{\Delta} = 3/2$ and that this is lower than the dimension $\Delta$ of any bosonic operator $\cO$ which one can construct. In particular, the lowest dimension bosonic operator which is flavour-antisymmetric is $\cO = \Phi_{[i}^\dagger \partial \Phi_{j]}^\dagger$ with $\Delta = 7/2$.

There is a nice interpretation of this failure of the duality: there are \textit{too many flavours} for the bosonic theory to support a sufficiently low-dimension operator to match the fermionic theory. This fits in with a general expectation: bosonization works whenever $N_f \le N$, but not for $N_f > N$ \cite{hsin}.

Let us try and generalize this issue. Suppose we consider instead $U(1)$ bosons with $k \ge 2$, so that now the fermions carry an $SU(k)$ spin but sit at level $1$.
\begin{align*}
N_f=1:&& \cO &= \Phi^{\dagger} \Phi^{\dagger} & \tilde{\cO} &= \Psi^{\dagger [a} \Psi^{\dagger b]} & \Delta &= \tilde{\Delta} = 2 + \frac{1}{3} = \frac{7}{3}  \\
N_f \ge 2:&& \cO_S &= \Phi_{(i}^{\dagger}\Phi_{j)}^{\dagger} & \tilde{\cO}_S &= \Psi_{(i}^{\dagger [a} \Psi_{j)}^{\dagger b]} & \Delta &= \tilde{\Delta} = 2 + \frac{1}{3} = \frac{7}{3} \\
&&\cO_A &= \Phi_{[i}^{\dagger } \partial \Phi_{j]}^{\dagger} & \tilde{\cO}_A &= \Psi_{[i}^{\dagger [a} \partial\Psi_{j]}^{\dagger b]} & \Delta &= \tilde{\Delta} = 2 + 1 + \frac{1}{3} = \frac{10}{3}
\end{align*}
Here we have listed the lowest dimensional two-particle operators in both the symmetric and antisymmetric representations of $SU(N_f)$.
The striking this here is that $\Psi_{[i}^{\dagger (a} \Psi_{j]}^{\dagger b)}$ is not a valid contender for $\tilde{\cO}_A$, since it transforms in a non-integrable representation of $SU(2)$ at level 1. (This operator would have $\tilde{\Delta} = 5/3$.) Thus it seems like the problem has gone away.

We could try a different tack and consider instead $U(2)$ bosons with $k=1$, again with $2$ particles. Similarly to the above, we are forbidden bosonic states which are gauge symmetric, so two particles must always be in a gauge singlet. This means that the lowest dimension operators look a little different according to whether $N_f \ge 2$ or not:
\begin{align*}
N_f=1:&& \cO &= \Phi^{\dagger [a} \partial \Phi^{\dagger b]} & \tilde{\cO} &= \Psi^\dagger \partial \Psi^\dagger & \Delta &= \tilde{\Delta} = 2 + 1 - \frac{1}{3} = \frac{8}{3} \\
N_f \ge 2:&& \cO_S &= \Phi_{(i}^{\dagger [a } \partial \Phi_{j)}^{\dagger b]} & \tilde{\cO}_S &= \Psi_{(i}^\dagger \partial \Psi_{j)}^\dagger & \Delta &= \tilde{\Delta} = 2 + 1 - \frac{1}{3} = \frac{8}{3} \\
&& \cO_A &= \Phi_{[i}^{\dagger [a }\Phi_{j]}^{\dagger b]} & \tilde{\cO}_A &= \Psi_{[i}^\dagger \Psi_{j]}^\dagger & \Delta &= \tilde{\Delta} = 2 - \frac{1}{3} = \frac{5}{3}
\end{align*}
But again, it seems as if there is no problem for arbitrarily large $N_f$.

Of course, we have only investigated the two-particle sector. It turns out that this is not usually sufficient to explore the issue we are interested in.

\subsubsection*{Three Particle Operators}

Consider $U(2)$ with $k=1$ and look at $p=3$ particle operators. We have already discussed $N_f = 1$ in the context of the indices, and seen that everything works out nicely once we understand how to implement the fusion rules correctly.

\def\Oyng(#1){{\cO_{\Yboxdim6pt \yng(#1)}}}
\def\Otyng(#1){{\tilde{\cO}_{\Yboxdim6pt \yng(#1)}}}
Now let us stick with $N=2$ and $k=1$ but move to $N_f = 2$ and then $N_f \ge 3$. We expect that the former should work well but that the latter should break. Here are the lowest-dimensional operators transforming in the specified (unreduced) $SU(N_f)$ representations:
\begin{align*}
N_f=2:&& \Oyng(3) &= \Phi^{\dagger [a}_{(i} \partial\Phi^{\dagger b]}_{j} \partial\Phi^{\dagger c}_{k)} & \Otyng(3) &= \Psi^\dagger_{(i} \partial\Psi^\dagger_{j} \partial^2 \Psi^\dagger_{k)} & \Delta &= \tilde{\Delta} = 5 \\
&& \Oyng(2,1) &= \Phi^{\dagger [a}_{[i} \Phi^{\dagger b]}_{j]} \Phi^{\dagger c}_k & \Otyng(2,1) &= \Psi^\dagger_{[i}\Psi^\dagger_{j]} \partial \Psi^\dagger_{k} & \Delta &= \tilde{\Delta} = 3 \\
N_f \ge 3:&& \Oyng(3) &= \Phi^{\dagger [a}_{(i} \partial\Phi^{\dagger b]}_{j} \partial\Phi^{\dagger c}_{k)} & \Otyng(3) &= \Psi^\dagger_{(i} \partial\Psi^\dagger_{j} \partial^2 \Psi^\dagger_{k)} & \Delta &= \tilde{\Delta} = 5 \\
&& \Oyng(2,1) &= \Phi^{\dagger [a}_{[i} \Phi^{\dagger b]}_{j]} \Phi^{\dagger c}_k & \Otyng(2,1) &= \Psi^\dagger_{[i}\Psi^\dagger_{j]} \partial \Psi^\dagger_{k} & \Delta &= \tilde{\Delta} = 3 \\
&& \Oyng(1,1,1) &= \Phi^{\dagger [a}_{[i} \Phi^{\dagger b]}_{j} \partial\Phi^{\dagger c}_{k]} & \Otyng(1,1,1) &= \Psi^\dagger_{[i}\Psi^\dagger_{j} \Psi^\dagger_{k]} & \Delta &= 4 \neq \tilde{\Delta} = 2
\end{align*}
Here we see that, indeed, for $N_f \ge 3$ the duality fails.

What about the original generalization we attempted, with $N = 1$, $k=2$ and $N_f = 2$? With three particles, we find that the bosonic state of minimal dimension in the $\Yboxdim6pt \Yvcentermath1 \yng(2,1)$ representation is
\begin{equation*}
\Oyng(2,1) = \Phi^{\dagger}_{[i} \partial\Phi^{\dagger }_{j]} \Phi^{\dagger}_k \qquad \text{with} \qquad \Delta = 5
\end{equation*}
which is dual in our sense to 
\begin{equation*}
\Otyng(2,1)' = \Psi^{\dagger [a}_{[i} \partial\Psi^{\dagger b]}_{j]} \partial\Psi^{\dagger c}_k
\end{equation*}
whereas in fact the lowest-dimension fermionic operator can be projected out from
\begin{equation*}
\Otyng(2,1) = \Psi^{\dagger [a}_{(i} \Psi^{\dagger b]}_{j)} \partial\Psi^{\dagger c}_k \qquad \text{with} \qquad \tilde{\Delta} = 4
\end{equation*}
showing that, indeed, the duality is still violated.

\subsection{$N_f > N$ Versus $N_f = N$}\label{nfnearn-sec}

We can easily find a violation by considering $N_f \ge N+1$ for arbitrary $k,N$, and looking at the lowest-dimensional operators transforming in the
\vspace{1.8em}\begin{equation*}
{\tikzmark{ystart}\Yautoscale0\Yboxdim19pt\yng(7,1,1,1)\tikzmark{yend}}
\begin{tikzpicture}[overlay, remember picture]%
\draw[decoration={brace,amplitude=7},decorate,thick] ($(ystart)+(0,2.8)$) -- node[above = 0.3cm] {$k$} ($(yend)+(-0.05,2.8)$);%
\draw[decoration={brace,amplitude=7},decorate,thick] ($(ystart)+(-0.2,0)$) -- node[left = 0.3cm] {$N+1$} ($(ystart)+(-0.2,2.6)$);%
\end{tikzpicture}%
\end{equation*}%
\cpttikzundefinethesenodes{ystart,yend}%
flavour representation (where we have included the singlets to indicate this is an object made of $k+N$ operators).

Let us set aside the special case $k=1$. (This is slightly different because there are no fusion rules in Abelian theories, but is straightforward to handle as in the examples above, and results in a \textit{larger} discrepancy.) Then the corresponding minimal bosonic and fermionic operators can be projected out from 
\begin{align*}
\cO &= \Phi^{\dagger [a_1}_{[i_1} \cdots \Phi^{\dagger a_N]}_{i_N]} \cdot \Phi^{\dagger (a_{N+1}}_{(i_{N+1}} \cdots \Phi^{\dagger a_{N+k-1}}_{i_{N+k-1}} \partial\Phi^{\dagger a_{N+k})}_{i_{N+k})} \\
\tilde{\cO} &= \Psi^{\dagger [a_1}_{(i_1} \cdots \Psi^{\dagger a_k]}_{i_k)} \cdot \Psi^{\dagger (a_{k+1}}_{[i_{k+1}} \cdots \Psi^{\dagger a_{k+N-1}}_{i_{k+N-1}} \partial\Psi^{\dagger a_{k+N})}_{i_{k+N}]}
\end{align*}
which indeed have different dimensions. To be concrete, their gauge representations and dimensions are, for the bosonic operator,
\vspace{1.5em}\begin{align*}
\cO:&\hskip7em{\tikzmark{ystart}\Yvcentermath1\Yautoscale0\Yboxdim19pt\yng(8,1,1)\tikzmark{yend}} \\
&\implies \Delta = (N+k) + 1 + \frac{k(k+1)-N(N-1)}{2(N+k)} = \frac{N+3k+3}{2}
\begin{tikzpicture}[overlay, remember picture]%
\draw[decoration={brace,amplitude=7},decorate,thick] ($(ystart)+(0,1.2)$) -- node[above = 0.3cm] {$k+1$} ($(yend)+(-0.05,1.2)$);%
\draw[decoration={brace,amplitude=7},decorate,thick] ($(ystart)+(-0.2,-0.9)$) -- node[left = 0.3cm] {$N$} ($(ystart)+(-0.2,1.1)$);%
\end{tikzpicture}%
\end{align*}
and for the fermionic operator
\vspace{1.4em}\begin{align*}
\tilde{\cO}:& \hskip7em {\tikzmark{ystart2}\Yvcentermath1\Yautoscale0\Yboxdim19pt\yng(4,1,1,1,1,1,1)\tikzmark{yend2}} \\
&\implies \tilde{\Delta} = (N+k) + 1 - \frac{N(N+1)-k(k-1)}{2(N+k)} = \frac{N+3k+1}{2}
\begin{tikzpicture}[overlay, remember picture]%
\draw[decoration={brace,amplitude=7},decorate,thick] ($(ystart2)+(0,2.5)$) -- node[above = 0.3cm] {$N+1$} ($(yend2)+(-0.05,2.5)$);%
\draw[decoration={brace,amplitude=7},decorate,thick] ($(ystart2)+(-0.2,-2.2)$) -- node[left = 0.3cm] {$k$} ($(ystart2)+(-0.2,2.4)$);%
\end{tikzpicture}%
\end{align*}
\cpttikzundefinethesenodes{ystart,yend,ystart2,yend2}%
so that in particular
\begin{equation*}
\tilde{\Delta} = \Delta - 1 \fstp
\end{equation*}

At a mechanical level, the key feature of having $N_f > N$ is that the bosonic theory is forced into using a derivative and gauge symmetrization to support the large flavour antisymmetrization. On the other hand, the fermion can take advantage of the large flavour antisymmetrization to keep a relatively low dimension.





\subsubsection*{Setting $N_f = N$}

By contrast, at the special point $N_f = N$ everything is very nice. When the number of flavours and colours within the bosonic theory coincide in this way, the lowest-dimension gauge-singlet operators take the form
\begin{equation}
\cO_r = \left(\Phi^{[a_1}_{[i_1} \cdots \Phi^{a_N]}_{i_N]}\right)^{r}
\end{equation}
which have the property that in $SU(N)_k$ their dimension scales linearly with $r$,
\begin{equation}
\Delta = \frac{(1+2kN+N^2)r}{2(N+k)} \fstp
\end{equation}
Their angular momentum is $\cJ = 0$ -- they are non-interacting, spinless bosons, decoupling completely from the $SU(N)$ Chern-Simons theory save for the constant shift to their dimension.

The operators $\cO_r$ are dual to fermionic operators roughly of the form
\begin{equation}
\tilde{\cO}_r = \left(\Psi^{(a_1}_{[i_1} \cdots \Psi^{a_N)}_{i_N]}\right)^{k} \left(\partial\Psi^{(b_1}_{[j_1} \cdots \partial\Psi^{b_N)}_{j_N]}\right)^{k} \cdots \left(\partial^{\lfloor r/k \rfloor}\Psi^{(c_1}_{[l_1} \cdots \partial^{\lfloor r/k \rfloor}\Psi^{c_N)}_{l_N]}\right)^{r \operatorname{mod} k}
\end{equation}
which are indeed also of matching minimal dimension in the $U(k)_{-N}$ theory. (Each $k$th power is understood as having a gauge antisymmetrization. Note that one cannot start symmetrizing the blocks of $N$ together until one has formed up these sets of $U(k)$ baryons.)

Notice that the lowest-dimension bosonic operators break
\begin{equation}
SU(N)_{\text{flavour}} \times SU(N)_{\text{colour}} \to SU(N)_\text{diagonal}
\end{equation}
in exactly the same way that BPS vortices do in the phase of this theory obtained by adding a chemical potential. (See \cite{cpt-susyqhe,cpt-nastates,cpt-dualityqhe} for a discussion of this theory from the point of view of quantum Hall physics.)

One reason that this is a particularly nice point is that there is, of course, a \textit{constrained fermion} realization of the affine Lie algebra  of $SU(k)_{-N}$ consisting of a theory with $N_f = N$ flavours of fermion \cite{naculich}. Our chiral states are in bijection with constrained fermion states, though in our picture their dimension is not due to a quartic rearrangement of their stress-energy tensor (via the Sugawara construction) but due to the flux attachment, and the properties of the attached Chern-Simons Wilson lines. The difference in the approach is clear from the fact that $N_f$ and $N$ are independent parameters for us.

Nonetheless, it is clear that $N_f = N$ is on algebraic grounds a very natural particular case of the bosonization duality. Moreover, we note that all the $N_f < N$ dualities can of course be obtained from the $N_f = N$ duality by simply restricting to sectors where one or more flavours of boson or fermion are not turned on. In this sense, the $N_f = N$ case is the natural parent for all these bosonization dualities.\footnote{If one prefers to think in terms of integrating out fields, note one can choose to either integrate out fermions and bosons using one of two sign choices for the `mass' term. One choice simply removes a flavour on both sides of the theory. The other choice additionally shifts the Chern-Simons level $-N \to -(N-1)$ on the fermionic side and partially Higgses the gauge field on the bosonic side, breaking $SU(N) \to SU(N-1)$.}

It would be nice to understand the whole picture of bosonization (beyond chiral states) more algebraically. It seems likely that attacking the $N_f = N$ duality is the right way to do this.

\subsection{Unitarity Violations: Good, Bad, Ugly}

We have seen that when the number of flavours $N_f$ is larger than $N$, the bosonization duality necessarily fails. One perspective on the mechanical reason for this was that the large number of flavours allows us to construct fermionic operators which are suitably antisymmetrized without using many derivatives; intuitively speaking, the fermions can get too close. The dimension of the resulting operator is too low to be dual to any bosonic operator.

There is a natural concern which might arise: can the fermions get \textit{too} close? Specifically, there is a unitarity bound \cite{nagain} in these theories given by
\begin{equation}
\Delta, \tilde{\Delta} \ge 1
\end{equation}
which is saturated only by free operators. Violations of this bound for $k<0$ were discussed in \cite{cpt-anyons,cpt-nonanyon} in the context of the bosonic theory, where they coincided with the appearance of Jackiw-Pi vortices (a non-topological soliton which solves the classical equation of motion) in the spectrum.

For our better behaved $k>0$ case, one can reasonably ask two questions. Firstly, do we ever fall foul of the unitarity bound? Secondly, does this have anything to do with the violations of the duality for large numbers of flavours?

The answer to both questions is yes. In this section, we will find operators which cause unitarity problems precisely for $N_f > N$.

This is reminiscent of similar situations in superconformal field theories in which the relationship between $N_f$ and $N$ affects the IR fate of theories \cite{seiberg-emduality,gaiotto-witten-bc,asselbad} -- in particular, both whether or not the naive implementation of superconformal symmetry leads to unitarity violations (which mean there are additional sectors of free operators), and whether or not certain Seiberg-like dualities hold. We will comment on this at the end of this section.

\subsubsection*{Abelian Fermionic Theories, $k=1$}

Here, although we have focussed on $k > 0$, it appears that including a large number of fermion flavours can drive the dimension down towards the unitarity bound. Indeed, setting $k=1$, consider the Abelian fermionic operator
\begin{equation}
\tilde{\cO} = \Psi^\dagger_{[i_1} \Psi^\dagger_{i_2} \cdots \Psi^\dagger_{i_{N_f}]} \fstp
\end{equation}
The dimension of this operator, in $U(1)_\ell$, is
\begin{equation}
\tilde{\Delta} = N_f - \frac{N_f(N_f-1)}{2 \ell} \fstp
\end{equation}
Then we find that 
\begin{equation}
\tilde{\Delta} \ge 1 \qquad \iff \qquad N_f \le 2\ell
\end{equation}
and in particular the bound is violated for $N_f$ larger than this. It seems that a free operator arises at $N_f = 2\ell$, and then must decouple from the theory as $N_f$ grows beyond this point.

But this is not necessarily the worst-behaved operator. Consider the sequence
\begin{equation}
\tilde{\cO}_q = \left(\Psi^\dagger_{[i_1} \cdots \Psi^\dagger_{i_{N_f}]}\right) \left(\partial\Psi^\dagger_{[i_1} \cdots \partial\Psi^\dagger_{i_{N_f}]}\right) \cdots \left(\partial^{q-1}\Psi^\dagger_{[i_1} \cdots \partial^{q-1}\Psi^\dagger_{i_{N_f}]}\right)
\end{equation}
instead. Then we find that
\begin{equation}
\tilde{\Delta} = qN_f + \frac{1}{2} N_f q(q-1) - \frac{qN_f(qN_f-1)}{2\ell} \ge 1 \qquad \forall q \qquad \iff \qquad N_f \le \ell \fstp
\end{equation}
More specifically, at the value $N_f = \ell$, this sequence has $\tilde{\Delta} \propto q$, whilst away from this point the quadratic dependence on $q$ has different signs for $N_f \gtrless \ell$. Therefore, for any $N_f > \ell$ there are infinitely many operators (for sufficiently large $q$) which naively have increasingly negative dimensions.

The conclusion is that these operators are not being handled correctly. As described in \cite{cpt-anyons,cpt-nonanyon}, they correspond to non-normalizable states under the state-operator map, and should be removed from the spectrum. But it is tempting to speculate that perhaps we should also include some free, decoupled operators in the theory corresponding to all these illegal states, much like the Jackiw-Pi vortices did in the bosonic $k < 0$ case. We will not pursue such speculation here.

Given we can adjust the effective $U(1)$ level $\ell$ in the dualities, via the extra $U(1)_n$ factor from the introduction, though, it is not immediately clear that this phenomenon can be at all related to the duality.

However, suppose that we compute the dimension of the above operators for general $n,N$. Then the effective value of the statistical parameter $1/\ell$ is
\begin{equation}
\frac{1}{\ell} = \frac{n}{nN+1}
\end{equation}
Notice that if we focus on $n \ge 0$ (an additional constraint which we think of as analogous to $k>0$) then this quantity ranges between $0$ and $1/N$. It follows that, if we want $\tilde{\Delta} \ge 1$ for all $q,n\ge 0$ then $N_f \le N$. One reaches the same conclusion if one requires only that the operators do not violate the bound at $n=\infty$, where the duality is $SU(N)_k \leftrightarrow U(k)_{-N}$.

(Note that for $n < 0$, it is much easier for operators to violate the unitarity bound; this is more reminiscent of the situation with Jackiw-Pi vortices, for bosonic theories with $k < 0$. We will not be interested in these sorts of violations.)

This suggests that perhaps there is indeed a link between violations of the unitarity bound (for either positive $n$ or simply the particular case $n=\infty$) and the duality.

\subsubsection*{Non-Abelian Fermionic Theories, $k>1$}

The above generalizes fairly straightforwardly to $U(k)$ fermionic theories. If we consider the combination of $q N_f$ baryons as a simple probe, and set $n=\infty$ for simplicity, we find a lowest dimension of
\begin{align*}
\tilde{\Delta} = q N_f k + \frac{1}{2}q(q-1) N_f k - \frac{kq N_f(q N_f-1) - qN_F k(k-1)}{2(N+k)} \\
 - \frac{qN_f k(qN_f k-1)}{2} \left[ \frac{1}{k N} - \frac{1}{k(N+k)} \right]
\end{align*}
and then because of the quadratic terms, once more
\begin{equation}
\tilde{\Delta} \ge 1 \quad \forall q \qquad \iff \qquad N_f \le N \fstp
\end{equation}
Once again, the boundary of validity of bosonization coincides with a unitarity bound.

Notice that, on the bosonic side, we do not get these violations of the unitary bound for our $k>0$ theories. The basic reason is easy to understand, focussing on $SU(N)_k$ for definiteness. Only gauge antisymmetrizations contribute negatively to the anomalous dimension $\Delta$, but we cannot construct operators with arbitrarily large gauge antisymmetrizations. In particular, we cannot construct large operators with quadratically negative contributions to the anomalous dimension. Actually, for a single baryon (which is as bad as it gets) we see
\begin{equation}
\cO = \Phi^{\dagger[a_1}_{[i_1} \cdots \Phi^{\dagger a_N]}_{i_N]} \qquad \implies \qquad \Delta = N - \frac{N^2-1}{2(k+N)} > 1 \fstp
\end{equation}

\subsubsection*{Good, Bad, Ugly}

As we pointed out above, this picture is somewhat reminiscent of the story of ``the good, the bad and the ugly'' \cite{gaiotto-witten-bc,asselbad}. Here, there is a Seiberg-like duality between two superconformal theories which works only for certain values of $N_f$ relative to $N$. Moreover, this coincides with whether the unitarity bound in the theory is (naively) violated or not.

The ``good'' theories are those for which $N_f \ge 2N$, which interestingly is for sufficiently \textit{large} numbers of flavours. Here, the naive implementation of the unitarity bound in the IR leads to no problems, and we also find 3d ${\cal N} = 4$ SQCD infrared dualities of the form
\begin{equation*}
U(N) + N_f \text{ hypermultiplets} \quad \longleftrightarrow \quad U(N_f - N) + N_f \text{ hypermultiplets} \fstp
\end{equation*}
Meanwhile, a borderline ``ugly'' case at $N_f = 2N-1$ saturates the unitarity bound, which slightly deforms the duality to
\begin{align*}
U(N) + (2N-1) \text{ hypermultiplets} \quad \longleftrightarrow \quad U(N-1) + (2N-1) \text{ hypermultiplets} \\ + 1 \text{ free hypermultiplet} \fstp
\end{align*}
For $N_f < 2N-1$, outright violations of the bound occur, and no simple infrared duality like the above holds. (In \cite{asselbad}, a suggestion \cite{yaakov} for a duality in the ``bad'' cases $N \le N_f < 2N-1$ paralleling the ``ugly'' case was analyzed, and shown to miss some important global features of the IR theories.)

Despite being for a somewhat different parameter range, this clearly resembles what we have seen for our (``good'') $N_f \le N$ theories and (``bad'') $N_f > N$ theories. Our proposed fermionic dual is clearly somewhat subtle for  large $N_f$ in that, proceeding naively, we can construct various disallowed operators. It is no surprise that it is in fact not the dual of the bosonic theory, which is well-behaved in this respect. 

Nonetheless, the fermionic theory does contain a \textit{subsystem} of operators which \textit{are} dual to the bosonic operators.  (This is reminiscent of the partially successful ``bad'' duals alluded to above.) This leaves open the possibility that there is some duality of $U(N)_k$ theory with $N_f$ bosons to a theory of fermions subject to some additional constraint which projects out the unwanted states; perhaps with the structure of a coset model.

It would be interesting to explore this further.


\clearpage
\section{Conclusion}\label{conclusion}

We have described the non-relativistic bosonization duality
\begin{equation*}
U(N)_{k,k+nN} + N_f \text{ fund. scalars} \quad \longleftrightarrow \quad U(k)_{-N} \times U(1)_n + N_f \text{ fund. fermions}
\end{equation*}
for $N_f \le N$, and how it fails for $N_f > N$, by matching individual chiral operators in these theories, and seeing how the corresponding indices also agree. It would be nice to extend this to non-chiral operators, for which explicit dimensions are not known, to try to gain insight about their spectrum.

We also noted that $N_f = N$ is a particularly natural point to study, since e.g. there is a free fermion representation of the affine Lie algebra $su(k)_N$ in terms of $N$ fermions, and since the other dualities follow from this one upon removing some flavours. It would be interesting to try and understand the full spectrum at this point in terms of the affine Lie algebra.

We have seen that the bosonization fails precisely for the same theories as a unitarity bound on operator dimension is violated, in analogy with the story of supersymmetric Seiberg-like dualities (which require instead \textit{lower bounds} like $N_f \ge 2N$). We have also speculated that, if a mechanism which restricts which states are allowed in the fermionic theory is added, it might be possible to find a duality which does work for larger flavours. Perhaps this might take the form of a Seiberg duality whose supersymmetry has been broken.

One lesson offered by this work is that since non-relativistic field theories can be controlled as well as supersymmetric theories, with very little complication or excessive field content, they provide a simplified laboratory in which e.g. non-supersymmetric dualities can be probed. We emphasize, for example, that the restriction to fundamental fields in this work is not necessary; the conformal dimensions for states built from operators in \textit{arbitrary} representations are known (and are presented in the context of the relevant field theories in \cite{cpt-nonanyon}). It would clearly be interesting to use this approach to conjecture and test more dualities.


\acknowledgments

Thanks to Alec Barns-Graham for helpful discussions, and to Nima Doroud, David Tong and Nick Dorey for comments on the manuscript. CPT is supported by a Junior Research Fellowship at Gonville \& Caius College, Cambridge; much of the work was also completed whilst supported by the European Research Council under the European
Union's Seventh Framework Programme (FP7/2007-2013), ERC grant
agreement STG 279943, ``Strongly Coupled Systems". CPT is also grateful for hospitality from National Taiwan University whilst the work was being finalized. This work has been partially supported by STFC consolidated grant ST/P000681/1.


\clearpage
\bibliography{scbosonization}{}

\end{document}

%% file: Map.pdf_tex
\begingroup%
  \makeatletter%
  \providecommand\color[2][]{%
    \errmessage{(Inkscape) Color is used for the text in Inkscape, but the package 'color.sty' is not loaded}%
    \renewcommand\color[2][]{}%
  }%
  \providecommand\transparent[1]{%
    \errmessage{(Inkscape) Transparency is used (non-zero) for the text in Inkscape, but the package 'transparent.sty' is not loaded}%
    \renewcommand\transparent[1]{}%
  }%
  \providecommand\rotatebox[2]{#2}%
  \ifx\svgwidth\undefined%
    \setlength{\unitlength}{445.4674476bp}%
    \ifx\svgscale\undefined%
      \relax%
    \else%
      \setlength{\unitlength}{\unitlength * \real{\svgscale}}%
    \fi%
  \else%
    \setlength{\unitlength}{\svgwidth}%
  \fi%
  \global\let\svgwidth\undefined%
  \global\let\svgscale\undefined%
  \makeatother%
  \begin{picture}(1,0.6585225)%
    \put(0,0){\includegraphics[width=\unitlength,page=1]{Map.pdf}}%
    \put(0.65821102,0.14359061){\color[rgb]{0,0,0}\makebox(0,0)[b]{\smash{$I=1$}}}%
    \put(0.7592285,0.14359061){\color[rgb]{0,0,0}\makebox(0,0)[b]{\smash{$I=2$}}}%
    \put(0.86024598,0.14359061){\color[rgb]{0,0,0}\makebox(0,0)[b]{\smash{$I=3$}}}%
    \put(0.96126346,0.14359061){\color[rgb]{0,0,0}\makebox(0,0)[b]{\smash{$I=4$}}}%
    \put(0,0){\includegraphics[width=\unitlength,page=2]{Map.pdf}}%
    \put(0.0521062,0.42138869){\color[rgb]{0,0,0}\makebox(0,0)[b]{\smash{$I=1$}}}%
    \put(0.0521062,0.26986246){\color[rgb]{0,0,0}\makebox(0,0)[b]{\smash{$I=2$}}}%
    \put(0.0521062,0.11833619){\color[rgb]{0,0,0}\makebox(0,0)[b]{\smash{$I=3$}}}%
    \put(0.0521062,0.01731871){\color[rgb]{0,0,0}\makebox(0,0)[b]{\smash{$I=4$}}}%
    \put(0,0){\includegraphics[width=\unitlength,page=3]{Map.pdf}}%
  \end{picture}%
\endgroup%

%% file: Map-SU.pdf_tex
\begingroup%
  \makeatletter%
  \providecommand\color[2][]{%
    \errmessage{(Inkscape) Color is used for the text in Inkscape, but the package 'color.sty' is not loaded}%
    \renewcommand\color[2][]{}%
  }%
  \providecommand\transparent[1]{%
    \errmessage{(Inkscape) Transparency is used (non-zero) for the text in Inkscape, but the package 'transparent.sty' is not loaded}%
    \renewcommand\transparent[1]{}%
  }%
  \providecommand\rotatebox[2]{#2}%
  \ifx\svgwidth\undefined%
    \setlength{\unitlength}{394.60031584bp}%
    \ifx\svgscale\undefined%
      \relax%
    \else%
      \setlength{\unitlength}{\unitlength * \real{\svgscale}}%
    \fi%
  \else%
    \setlength{\unitlength}{\svgwidth}%
  \fi%
  \global\let\svgwidth\undefined%
  \global\let\svgscale\undefined%
  \makeatother%
  \begin{picture}(1,0.20499956)%
    \put(0,0){\includegraphics[width=\unitlength,page=1]{Map-SU.pdf}}%
    \put(0.24472944,0.00632562){\color[rgb]{0,0,0}\makebox(0,0)[b]{\smash{Transpose}}}%
    \put(0.51896711,0.00632562){\color[rgb]{0,0,0}\makebox(0,0)[b]{\smash{$\sigma$}}}%
    \put(0.81356904,0.00632562){\color[rgb]{0,0,0}\makebox(0,0)[b]{\smash{$\sigma$}}}%
  \end{picture}%
\endgroup%